\documentclass[aps,prb,twocolumn,superscriptaddress,groupedaddress,footinbib]{revtex4-1} 

\usepackage{longtable}
\usepackage{morefloats}
\usepackage[dvips]{graphicx}
\usepackage{color}
\usepackage{epsfig,graphicx,amsfonts,amsbsy}
\usepackage{amsmath,amsfonts,amsthm,amssymb}
\usepackage{appendix}
\usepackage{makeidx}
\usepackage{url}
\usepackage{verbatim}
\usepackage[bookmarksnumbered,pdfpagelabels=true,plainpages=false,colorlinks=true,linkcolor=blue,citecolor=red,urlcolor=blue]{hyperref}
\usepackage[rightcaption]{sidecap}
\usepackage{array}
\usepackage{multirow}
\usepackage{tabularx}
\usepackage{braket} 
\usepackage{dsfont}
\usepackage{bbold}
\usepackage{etoolbox}
\usepackage{comment}

\begin{document}

\title{Universal quantum operation of spin-3/2 Blume-Capel chains}

\author{Silas Hoffman}
\author{Yiyuan Chen}
\author{Hai-Ping Cheng}
\author{X.-G. Zhang}
\address{Department of Physics and the Quantum Theory Project,
University of Florida,
Gainesville, FL 32611}
\address{Center for Magnetic Molecular Quantum Materials, University of Florida, Gainesville, FL 32611}
\date{\today}
	
\begin{abstract}
We propose a logical qubit based on the Blume-Capel model: a higher spin generalization of the Ising chain and which allows for an on-site anisotropy preserving rotational invariance around the Ising axis. We show that such a spin-3/2 Blume-Capel model can also support localized Majorana bound states at the ends of the chain. Inspired by known braiding protocols of these Majorana bound states, upon appropriate manipulation of the system parameters, we demonstrate a set of universal gate operations which act on qubits encoded in the doubly degenerate ground states of the chain.
\end{abstract}
\maketitle

\section{Introduction}

A single spin-1/2 particle provides a natural two-level system to support a qubit. According to the Loss-DiVincenzo realization of such a qubit,\cite{lossPRA98} pulsed control of electric and magnetic fields enables one- and multi-qubit quantum gates which are sufficient for universal quantum computation. In this setup, and most other setups for large scale quantum computing, pulsing the controllable parameters imparts a dynamical phase which enables conventional gate control. Alternatively, adiabatic and cyclic control of the magnetic field imparts a geometric phase, i.e. a non-Abelian Berry's phase or a holonomy,\cite{wilczekPRL84} in addition to the dynamical phase. Exploiting the holonomies to perform a quantum operation is known as holonomic quantum computing.\cite{zanardiPRA99} As compared to dynamic gates, geometric gates benefit from tolerance to fluctuations.\cite{bergerPRA13} For a spin-1/2, the group generated by such holonomies is the Abelian U(1) group\cite{berryPRRL84} and is insufficient for universal quantum computation.

To achieve universal quantum computation the holonomy group must be non-Abelian and necessitates both (1) two or more degenerate states and (2) auxiliary states; these are known as dark and bright states, respectively, in the trapped ion literature.\cite{duanSCI01} Consequently, realization of a universal set of quantum gates generated by the holonomy necessitates a system with more than two levels. Although there are several systems in which geometric control of qubits has been realized, we restrict our interest to spin-based qubits.

One route to realizing this geometric control in spin systems is higher spin particles.\cite{zeePRA88} In particular, it is known that adiabatically controlling the anisotropy of spin-3/2 particles generates an SU(2) holonomy group.\cite{avronPRL88,avronCMP89} In this case, because the anisotropies guarantee time-reversal symmetry, there are two pairs of degenerate states; either of which can be used to furnish the physical qubit basis while the others are auxiliary and facilitate operation. These states can be realized as heavy holes in quantum dots whereby an electric field, which couples to the aniostropy via the Stark effect or the quadrapole moment of the field, can control the anisotropies and thereby perform any single-qubit quantum gate.\cite{bernevigPRL06,budichPRB12hol} Moreover, it has recently been shown that entanglement of such holes can be geometrically generated when they are simultaneously coupled to an electromagnetic cavity.\cite{wysokinskiCM20}

Rather than considering larger spins, auxiliary states can be provided by more spins as in, for instance, the Ising spin-1/2 chain. In such a system, it is convenient to express the spins as nonlocal fermions\cite{liebAoP61} which, for a chain of finite length, host Majorana bound states (MBSs) in the fermionic representation. As exchanging two MBSs generates a quantum operation\cite{ivanovPRL01}, an analogous manipulation of the Ising chain provides a nonuniversal set holonomic quantum gates\cite{backensPRB17} which must be aided by dynamic operations to realize universal quantum control. For instance, in Ref.~\onlinecite{tserkovnyakPRA11}, the authors considered a spin-1/2 Ising chain in which the accumulation of geometric phase, upon manipulating the direction of the anisototropic exchange interaction, was supplemented by an applied magnetic field to perform single-qubit quantum operations. A pulsed exchange interaction between spin chains imparted a dynamic phase, sufficient to entangle two qubits.

In an effort to realize an entirely geometric manipulation of a spin chain, we study a higher spin generalization of a spin-1/2 Ising chain known as the Blume-Capel model which includes an on-site anistropy along the Ising axis. Like the spin-1/2 Ising chain, we find that spin-3/2 chains with rotational symmetry also support MBSs localized to the ends of the chain. Moreover, we find that a single-site chain, i.e. an isolated spin-3/2 state, also host zero energy MBSs which can be effectively braided by adiabatic control of the on-site anistoropy. By exploiting this Majorana representation, we discover an entirely geometric protocol for single- and two-qubit operations using the spin-3/2 states. These protocols can be extended from qubits furnished by one spin-3/2 particle to chains of spin-3/2 particles, thus providing an entirely geometric protocol for universal quantum computation of high spin chains.

Our paper is organized as follows: in Sec.~\ref{chain}, we consider a chain of spin-3/2 states and show that, upon transforming to the fermionic description, MBSs reside at the ends and can be used to encode the ground states of our logical qubit. We continue by describing the procedure for quantum operation, including initialization and one- and two-qubit gate operations, in Sec.~\ref{quop}. We consider the necessary ingredients to extend our system to higher spins in Sec.~\ref{disc} and make some concluded remarks in Sec.~\ref{conc}.

\section{High spin model}

\label{chain}

Our starting point is the well-known Blume-Capel model\cite{blumePR66,*capelP66},
\begin{eqnarray}
H_\textrm{BC}=-J\sum_{i=1}^{N-1}S^z_iS^z_{i+1}-K\sum_{i=1}^N\left(S^z_i\right)^2\,,
\label{bcm}
\end{eqnarray}
which generalizes the Ising chain by including an on-site anisotropy, parameterized by $K$, in addition to anisotropic exchange, parameteried by $J$. As we show in this section, this model supports MBSs which furnish a natural basis to encode a single qubit. However, because the available parameters to manipulate the qubit are insufficient, we describe a sufficient set of parameters that further generalize  our spin model.

In Eq.~(\ref{bcm}), $S^z_i$ is the generator of rotation around the $z$ axis of the pin at site $i$ and, for concreteness, we focus on the $S=3/2$ case wherein
\begin{equation}
S^z=\left(
\begin{array}{cccc}
 3/2 & 0 & 0 & 0\\
0 & 1/2 & 0& 0\\
 0 &0 & -1/2 & 0\\
 0 & 0 & 0 & -3/2
\end{array}\right)\,.
\end{equation}
The spin-3/2 case is convenient because it can be understood in terms of Pauli matrices: one spin-3/2 operator can be mapped into a tensor product of two spin-1/2 generators of rotation, $S^z=2\mathbb1\otimes\sigma^z+4\sigma^z\otimes\mathbb 1$ or, for the chain, $S^z_i=\sigma^z_{2i-1}+4\sigma^z_{2i}$. Accordingly, Eq.~(\ref{bcm}) transforms from a chain of spin-3/2 particles to a ladder of spin-1/2 particles (Fig.~\ref{fig1}),
\begin{align}
 H&=-16K\sum_{i=1}^N\left(\sigma^z_{2i-1}\sigma^z_{2i}\right)-4J\sum_{i=1}^{N-1}\left(\sigma^z_{2i-1}\sigma^z_{2i+1}\right.\nonumber\\
 &\left.+2\sigma^z_{2i}\sigma^z_{2i+1}+2\sigma^z_{2i-1}\sigma^z_{2i+2}+4\sigma^z_{2i}\sigma^ z_{2i+2}\right)\,.
\label{lad}
\end{align}
\begin{figure}[!t]
\includegraphics[width=.7\columnwidth]{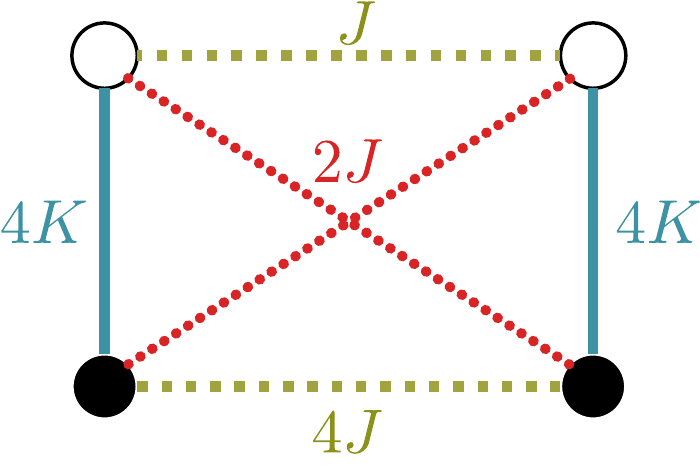}
\caption{A spin-3/2 Blume-Capel model with on-site anisotropy, $K$, and exchange interaction, $J$, can be mapped onto a ladder of exchange coupled spin-1/2, i.e. a quantum Ising model coupling spins up to three sites apart. The odd sites (white) couple to even sites (black) on the same rung with strength $4K$ and to sites one apart with strength $2J$. Odd sites couple to nearest neighbor odd (even) sites with strength $J$ ($4J$).}
\label{tex}
\label{fig1}
\end{figure}
This equation can be rewritten as an interacting spinless fermion system using the Jordan-Wigner mapping,\cite{liebAoP61} $\sigma_j^z=[\prod_{i<j}(1-2n_i)](c_j+c_j^\dagger)$ with $n_j=c_j^\dagger c_j$. Because each term in Eq.~(\ref{lad}) is, in general, a non-local product of Pauli matrices, they can be written as
\begin{align}
\sigma^z_i\sigma^z_j&=(c_i^\dagger-c_i)\left[\prod_{i< k<j}\left(1-2n_k\right)\right](c_j^\dagger+c_j)\,,
\end{align}
where we have assumed without loss of generality that $j>i$. We can further decompose the complex fermions into Majoranas according to the definition $c_j=(\gamma_j+i\gamma_j')/2$ wherin $c_j+c_j^\dagger=\gamma_j$, $c_j-c_j^\dagger=i \gamma_j'$,  and $1-2n_j=i\gamma_j\gamma_j'$. The products of Pauli matrices take the  form $\sigma^i_z\sigma^j_z=-i\gamma_i' \gamma_j\prod_{i< k<j}(i\gamma_k\gamma_k')$. Because the terms in Eq.~(\ref{lad}) follow this form with $1\leq i<j\leq 2N$, the first and last Majoranas, $\gamma_1$ and $\gamma_{2N}'$, are absent from the the Hamiltonian and are therefore zero energy operators that are localized to the ends of the chain; these are the MBSs. 

The degenerate ground states of this system are when all the spins point parallel or antiparallel to the $z$ axis which we denote by $|\Uparrow\rangle$ and $|\Downarrow\rangle$, respectively. The degenerate states can be characterized by the occupation of the zero energy complex fermion composed of the MBSs, $f=(\gamma_1+i\gamma_N')/2$. The physical meaning of these states is elucidated by noting that
\begin{align}
1-2f^\dagger f&=i\gamma_1\gamma_N'=\sigma_1^z\left[\prod_{j<N}\left(-\sigma_j^x\right)\right](-i\sigma_N^y)\nonumber\\
&=\sigma_1^z\left[\prod_{j\leq N}\left(-\sigma_j^x\right)\right](-\sigma_N^z)=\mathcal P\sigma^z_1\sigma^z_N\,,
\end{align}
with $\mathcal P$ a $\pi$ rotation at each site in the chain. That is, this operator transforms between the ground states: $(1-2f^\dagger f)|~\Uparrow~\rangle=|~\Downarrow~\rangle$ and $(1-2f^\dagger f)|~\Downarrow~\rangle=|~\Uparrow~\rangle$, i.e. the eigenstates of this operator are $|\pm\rangle=(|\Uparrow\rangle\pm|\Downarrow\rangle)/\sqrt{2}$ with $f^\dagger f|\pm\rangle=\pm|\pm\rangle$.

In order to operate our system, we require local control of the magnetic field transverse to the Ising axis, $H_B=-\sum_{i=1}^N(h_i S_i^x)$. In a further generalization to the Blume-Capel model, we suppose that the on-site anisotropy does not preserve the rotational symmetry around the $z$ axis but takes the rather general form\cite{avronPRL88}
\begin{equation}
H_A=-\sum_{a=1}^5 \left(d^a\Gamma^a\right)\,.
\label{Ha}
\end{equation}
Here, the anisotropies, quadratic in the spin-3/2 generators of rotation, can also be understood as tensor products of spin-1/2 matrices
\begin{align}
\Gamma^1&=\frac{1}{4\sqrt{3}}\left\{S^y_1,S^z_1\right\}=\sigma^z_1\sigma^x_2\,,\nonumber\\
\Gamma^2&=\frac{1}{4\sqrt{3}}\left\{S^z_1,S^x_1\right\}=\sigma^z_1\sigma^y_2\,,\nonumber\\
\Gamma^3&=\frac{1}{4\sqrt{3}}\left\{S^x_1,S^y_1\right\}=\sigma^y_1\,,\nonumber\\
\Gamma^4&=\frac{1}{4\sqrt{3}}[(S^x_1)^2-(S^y_1)^2]=\sigma^x_1\,,\nonumber\\
\Gamma^5&=\frac{1}{4}[(S^z_1)^2-(5/4)\mathbb1_{4\times4}]=\sigma^z_1\sigma^z_2\,,
\label{anis}
\end{align} 
where $d^a$ are the effective weights of the anisotropies. Because the anisotropy couples to the electric field through the Stark effect\cite{bernevigPRL06} or the quadrapole component\cite{budichPRB12hol}, we henceforth assume time-dependent control of $d^a$ which is essential to the operation of the qubit. Thus, the full combined Hamiltonian describing the system is $H=H_\textrm{BC}+H_A+H_B$. As we show below, we need only on-site anisotropy on the first so and so henceforth take $K=0$.

\section{Quantum Operation}
\label{quop}

In this section, we detail the initialization and operation of a qubit furnished by the states $|\pm\rangle$. Although our formulation of quantum operation is similar to the spin-1/2 Ising chain,\cite{tserkovnyakPRA11} we take advantage of the on-site anisotropy available in higher spin systems rather than relying on the anisotropy in the exchange interaction. 

\subsection{Operation of a spin-3/2 qubit}
\label{single}

To that end, we momentarily abandon the chain and focus on a single spin-3/2 site described by Eq.~(\ref{Ha}).  We remind the reader that because Eq.~(\ref{Ha}) preserves time reversal symmetry, Kramers theorem guarantees two degenerate states split by $\sqrt{\sum_a |d^a|^2}$ for any value of the five-vector $d^a$. By the same Jordan-Wigner transformation used in the previous section, $\Gamma^5=-i\gamma_1'\gamma_2$. Clearly in this case $[H_A,\gamma_1]=[H_A,\gamma_2']=0$, i.e. a single site hosts MBS, $\gamma_1$ and $\gamma_2'$, similarly to the chain. The four eigenstates of $S^z$, $|\uparrow\rangle$, $|\downarrow\rangle$, $|\upharpoonright\rangle$, $|\downharpoonright\rangle$, with eignenvalues $3/2$, $-3/2$, $1/2$, and $-1/2$, respectively, are simultaneously eigenstates of $\Gamma^5$. In particular, $\langle\Gamma^5\rangle=1$ ($\langle\Gamma^5\rangle=-1$) with the expectation values taken with respect to the state $|\uparrow\rangle$ or $|\downarrow\rangle$ ($|\upharpoonright\rangle$ and $|\downharpoonright\rangle$). The eigenstates of $i\gamma_1\gamma_2'$ with eigenvalues $\pm1$ are 
\begin{equation}
|\pm_{3/2}\rangle=(|\uparrow\rangle\pm|\downarrow\rangle)/\sqrt{2}
\end{equation} 
and 
\begin{equation}
|\pm_{1/2}\rangle=(|\upharpoonright\rangle\pm|\downharpoonright\rangle)/\sqrt{2}.
\end{equation}

\begin{figure}[!t]
\includegraphics[width=1\columnwidth]{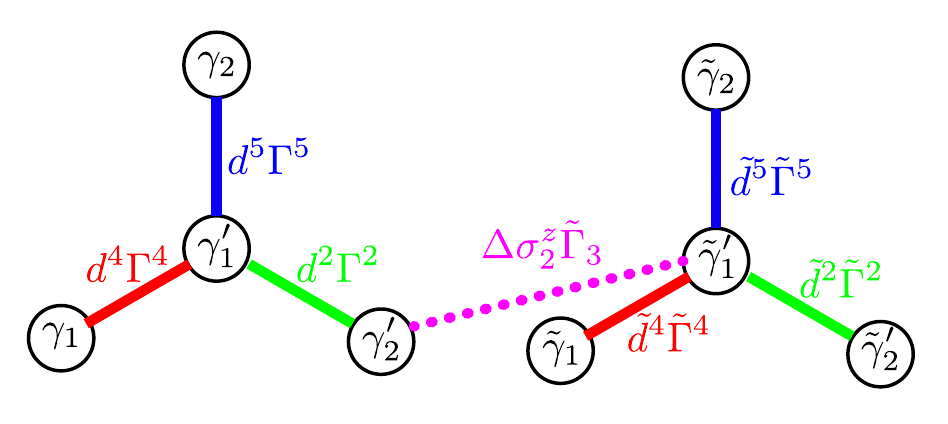}
\caption{Schematic of two spin-3/2 particles represented as Majorana fermions. Coupling of Majorana fermions within the same particle is controlled by the on-site anisotropy (highlighted in blue, green, and red) while the coupling between inter-spin Majorana fermions (highlighted in magenta) is controlled by a product of spin and anisotropy operators.}
\label{tex}
\label{spin_braid}
\end{figure}

\begin{figure}[!t]
\includegraphics[width=1\columnwidth]{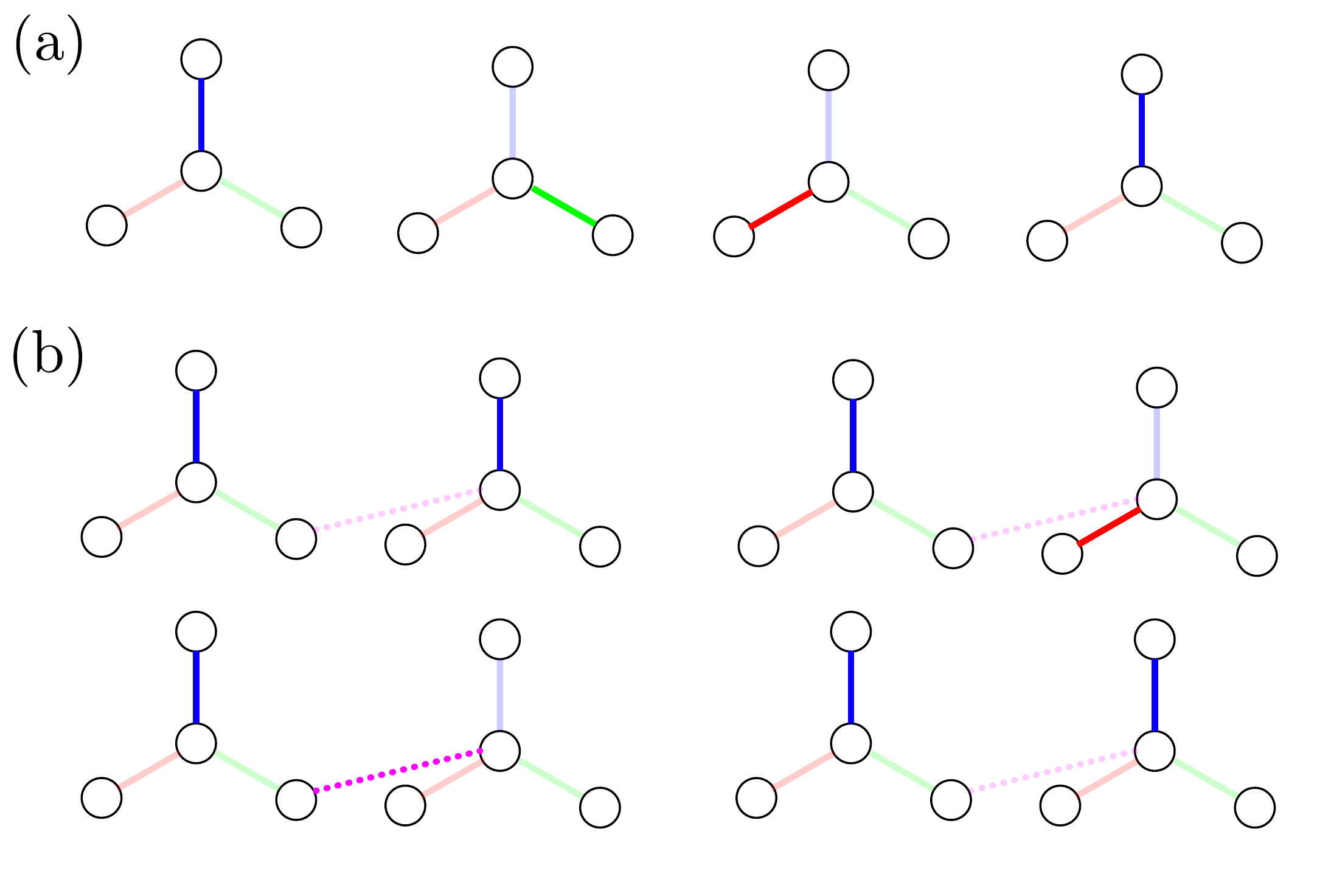}
\caption{Schematic of the necessary changes in the Hamiltonian parameters to braid Majorana fermions within (a) a single spin-3/2 qubit and (b) between spin-3/2 qubits.}
\label{tex}
\label{sb3}
\end{figure}

The other anisotropies [Eq.~(\ref{anis}) can, likewise, be fermionized to obtain
\begin{align}
\Gamma^1&=-i\gamma_1\gamma_2\gamma_2'\,,\,\,\Gamma^2=-i\gamma_1'\gamma_2'\,,\,\,\Gamma^3=\gamma_1'\,,\nonumber\\
\Gamma^4&=-i\gamma_1\gamma_1'\,,\,\,\Gamma^5=-i\gamma_1'\gamma_2\,,
\label{anis_MBS}
\end{align}
where we have repeated the transformed $\Gamma^5$ operator for completeness. Focusing on $\Gamma^2$, $\Gamma^4$, and $\Gamma^5$, the anisotropy has the structure of an inner MBS coupled to three outer MBSs, known in the literature as a $Y$-junction (Fig.~\ref{spin_braid})\cite{aliceaNATP11}. Borrowing the protocol from Ref.~\onlinecite{karzigPRX16}, one can braid two uncoupled MBSs: consider $d^1=d^3=0$ and parameterizing the remaining magnitudes of anisotropy by $\vec d=(d^2,d^4,d^5)=|\vec d|[\cos(\phi)\sin(\theta),\sin(\phi)\sin(\theta),\cos(\theta)]$. Consider an initial Hamiltonian with $\theta=\phi=0$, i.e. Eq.~(\ref{Ha}) with $d^5\neq0$ and the remaining $d^a=0$, and an initial state $|\psi\rangle=\alpha|+_{3/2}\rangle+\beta|-_{3/2}\rangle$ which is a ground state of that Hamiltonian. We proceed in three steps: (1) rotate $\vec d$ about the $y$ axis so that $\theta=0\rightarrow\theta=\pi/2$, (2) rotate $\vec d$ about the $z$ axis so that $\phi=0\rightarrow\phi=\varphi$, and (3) rotate $\vec d$ so that it once again points along the $z$ axis. $\hat d=\vec d/d$ traces out a solid angle on the unit sphere, $\Omega_Z=\varphi$ [Fig.~\ref{sb3}(a)]. This results in geometric phase of $\Omega_Z/2$ imprinted on the state: $|\psi\rangle\rightarrow|\psi\rangle=\exp( i\Omega_Z/2)\alpha|+_{3/2}\rangle+\exp( -i\Omega_Z/2)\beta|-_{3/2}\rangle$. That is, this operation corresponds to a rotation by angle $\Omega_Z$ around the $z$ axis of the Bloch sphere, $R_z(\Omega_Z)$, of a qubit defined on the basis $|\pm_{3/2}\rangle$. Although we have chosen a specific path that $\hat d$ traces out, the operation is independent of path for any solid angle traced out and rate-independent so long as the inverse time of operation is much smaller than $|\vec d|$. For $\Omega_Z=\pi/2$, this operation corresponds precisely to braiding MBSs $\gamma_1$ and $\gamma_2'$. The topological protection of this operation depends on the experimental realization of this model and can be guaranteed only so long as $d^a$ can be reasonably set to zero.\cite{karzigPRX16}

It is convenient to continue to use this MBSs picture when attempting to entangle states using a similar protocol. Consider an additional spin-3/2 site also described by Eq.~(\ref{Ha}); as a point of notation, we use tilde to distinguish parameters and operators of the second spin from the first spin. Because of the nonlocal string operator, the anisotropies on the second site, after the Jordan-Wigner transformation, take the following form:
\begin{align}
\tilde\Gamma^1&=i\gamma_1\gamma_1'\gamma_2\gamma_2'\tilde\gamma_1\tilde\gamma_2\tilde\gamma_2'\,,\,\,
\tilde\Gamma^2=-i\tilde\gamma_1'\tilde\gamma_2'\,,\nonumber\\
\tilde\Gamma^3&=-\gamma_1\gamma_1'\gamma_2\gamma_2'\tilde\gamma_1'\,,\,\,
\tilde\Gamma^4=-i\tilde\gamma_1\tilde\gamma_1'\,,\,\,
\tilde\Gamma^5=-i\tilde\gamma_1'\tilde\gamma_2\,.
\label{anis_MBS}
\end{align}
Because $\tilde \Gamma^2$, $\tilde \Gamma^4$, and $\tilde \Gamma^5$ take that same form as their untilded counterpart, it is clear that the single qubit operation $R_z(\Omega_Z)$ can be performed on the second spin-3/2. To entangle two spin states, one can braid two MBSs originating from different spins which requires a coupling between them. For instance, consider the Hamiltonian 
\begin{align}
H_\textrm{Ising}&=i\tilde\gamma_1'\vec\Delta\cdot\vec\gamma\,,\nonumber\\ 
\vec\Delta&=|\vec\Delta|[\cos(\phi)\sin(\theta),\sin(\phi)\sin(\theta),\cos(\theta)]\,,\nonumber\\
\vec\gamma&=(\gamma_2',\tilde\gamma_1,\tilde\gamma_2)\,.
\label{ising}
\end{align}
\begin{widetext}
Proceeding analogously to the single spin operation, we change the coupling so that the unit vector $\hat\Delta=\vec\Delta/|\vec\Delta|$ traces out a solid angle, $\Omega_\textrm{I}$ over the unit sphere [Fig.~\ref{sb3}(b)]. We find that this operates as Ising ZZ gate in the basis of $|\uparrow\rangle$ and $|\downarrow\rangle$,

\begin{equation}
\left(\begin{array}{c}
|\uparrow\rangle|\tilde\uparrow\rangle\\
|\uparrow\rangle|\tilde\downarrow\rangle\\
|\downarrow\rangle|\tilde\uparrow\rangle\\
|\downarrow\rangle|\tilde\downarrow\rangle
\end{array}\right)\rightarrow
\left(\begin{array}{cccc}e^{-i\Omega_\textrm{I}/2} & 0 &0&0\\
0 & e^{-i\Omega_\textrm{I}/2}  &0&0\\
0 & 0 &e^{-i\Omega_\textrm{I}/2} &0\\
0 & 0 &0&e^{i\Omega_\textrm{I}/2} \\
\end{array}\right)\left(\begin{array}{c}
|\uparrow\rangle|\tilde\uparrow\rangle\\
|\uparrow\rangle|\tilde\downarrow\rangle\\
|\downarrow\rangle|\tilde\uparrow\rangle\\
|\downarrow\rangle|\tilde\downarrow\rangle
\end{array}\right)\,,
\label{zz}
\end{equation}
or as an Ising XX gate in the $|\pm_{3/2}\rangle$-basis:
\begin{equation}
\left(\begin{array}{c}
|+_{3/2}\rangle|\tilde+_{3/2}\rangle\\
|+_{3/2}\rangle|\tilde-_{3/2}\rangle\\
|-_{3/2}\rangle|\tilde+_{3/2}\rangle\\
|-_{3/2}\rangle|\tilde-_{3/2}\rangle
\end{array}\right)\rightarrow
\left(\begin{array}{cccc}\cos(\Omega_\textrm{I}) & 0 &0&-i \sin(\Omega_\textrm{I})\\
0 & \cos(\Omega_\textrm{I})  &-i \sin(\Omega_\textrm{I})&0\\
0 & -i \sin(\Omega_\textrm{I}) &\cos(\Omega_\textrm{I}) &0\\
-i \sin(\Omega_\textrm{I}) & 0 &0&\cos(\Omega_\textrm{I}) \\
\end{array}\right)
\left(\begin{array}{c}
|+_{3/2}\rangle|\tilde+_{3/2}\rangle\\
|+_{3/2}\rangle|\tilde-_{3/2}\rangle\\
|-_{3/2}\rangle|\tilde+_{3/2}\rangle\\
|-_{3/2}\rangle|\tilde-_{3/2}\rangle
\end{array}\right)\,.
\label{xx}
\end{equation}

In order to couple the two spin-3/2 states in this way, we require a Hamiltonian which breaks time reversal symmetry, $i\tilde\gamma_1'\gamma_2'=\sigma_2^z\tilde\Gamma^3=[S_1^z-(S_1^z)^3]\tilde\Gamma^3$.
\end{widetext}

Although these operations are conveniently encoded in partially braiding MBSs, they are not sufficient for universal control of a qubit encoded by $|\pm_{3/2}\rangle$. Nonetheless, one may utilize Eq.~(\ref{Ha}) to encode employing the protocol introduced in Ref.~\onlinecite{budichPRB12}. In particular, to generate a rotation around the $y$ axis of the Bloch sphere by an angle $\varphi$, $d^3=d^5=0$ while
\begin{align}
d^2&=-\cos(\varphi)\sin(\Phi)\,,\nonumber\\
d^3&=\sin(2\varphi)\sin(\Phi/2)^2\,,\nonumber\\
d^5&=\sin(\varphi)^2+2\cos(2\varphi)\cos(\Phi)+2\cos(\Phi)\,.
\end{align}
Here, $\Phi$ is a function of time which changes from $0$ to $2\pi$. Again, because this is a geometric phase, the details of how $\Phi$ changes are unimportant so long as it changes adiabatically. Note that this does not have a convenient description in terms of MBSs as $\Gamma^3=\gamma_1'$. This operation with the single- and two- qubit operations described previously are sufficient for universal quantum computation.

Alternatively, one could redefine a single qubit to be two spin-3/2 particles in which a qubit is defined on the subspace $|\uparrow\rangle|\tilde\downarrow\rangle$ and $|\downarrow\rangle|\tilde\uparrow\rangle$. The above two-qubit protocol described above would allow a geometric navigation to any point on the Bloch sphere. A two-qubit protocol in the new basis could be achieved by coupling two spin-3/2 particles constituted in two different qubits.

\subsection{Initialization}
To initialize our system to $|+\rangle$, we begin with a large magnetic field on all the sites, $h_i/J\gg1$ with zero on-site anisotropy, $d^a=0$, so all the spins point along the positive $x$ axis, $|\Rightarrow\rangle=|\rightarrow\rightarrow\rightarrow\rangle$ with $S^x_i|\Rightarrow\rangle=(3/2)|\Rightarrow\rangle$. Upon decreasing the magnetic field on the first site, $h_1\rightarrow0$, and increasing the $z$-axis anisotropy , $d^5>0$, the first spin becomes a symmetric superposition of the spin aligned parallel and antiparallel to the $z$ axis, $(|\uparrow\rangle+|\downarrow\rangle)|\rightarrow\rightarrow\rangle/\sqrt{2}$. Note that because $h_i$ is large, $\langle S^z_2 \rangle=0$ so that the first site is effectively decoupled from the rest of the chain; in anticipation of subsequent operations, we note that this will be a convenient position in parameter space in which to perform gate operations. Upon decreasing the magnetic field on the second site, anisotropic exchange interaction aligns that site parallel to the first site, $(|\uparrow\uparrow\rangle+|\downarrow\downarrow\rangle)|\rightarrow\rangle/\sqrt{2}$. Proceeding analogously on subsequent sites, the system reaches the state $|+\rangle=(|\uparrow\uparrow\uparrow\rangle+|\downarrow\downarrow\downarrow\rangle)/\sqrt{2}$. Indeed, upon numerically simulating this manipulation of parameters, the state $\rightarrow\rangle$ which is initially orthogonal to $|+\rangle$, $\langle+|\psi(t_0)\rangle=0$, is manipulated into the state $|+\rangle$, $\langle+|\psi(t_f)\rangle=1$ (Fig.~\ref{fig1}).
\begin{figure}[!t]
\includegraphics[width=\columnwidth]{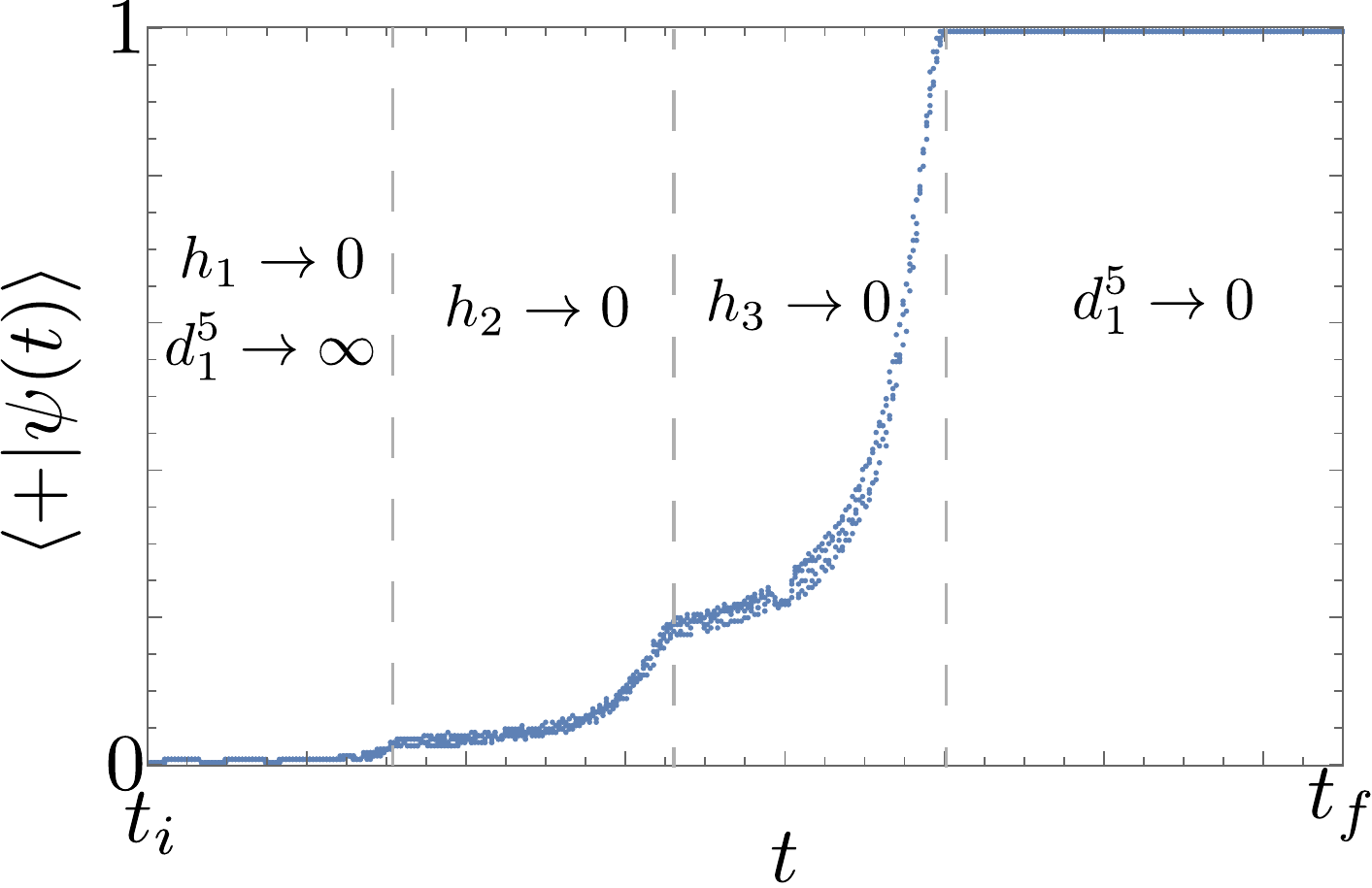}
\caption{Numerical simulation of our initialization procedure: beginning in a high magnetic field on all the sites, $h_i\gg K,J$, the anisotropy on the first sites is ramped up while the magnetic field ramped down. Ramping down the magnetic field on subsequent sites initializes the state into $|+\rangle=(|\Uparrow\rangle+|\Downarrow\rangle)/\sqrt{2}$.}
\label{tex}
\label{fig1}
\end{figure}

\subsection{Single-qubit gates}

Operation by a quantum gate can be, likewise, obtained by adiabatic control of our parameters. The essence of our operation relies on the ability to dynamically control the anisotropies of the first site to exploit the nonabelian holonomy of the spin-3/2 particle as in Sec.~\ref{single}. 

Consider a state that has been initialized to $|+\rangle$ and $d^a=h_i=0$. Similar to the initialization procedure, let us slowly ramp up the magnetic field to a large positive value on all the sites except the first site, $h_i\gg J$ for $i>1$. As the acquired Berry's phase is zero, the state adiabatically transforms to $(|\uparrow\rangle+|\downarrow\rangle)|\rightarrow\rightarrow\rangle/\sqrt{2}$. As the first site is now decoupled from the rest of the chain, we focus only on that state and, for notational convenience, only indicate the state of the first site. Exploiting the single qubit operations of a single spin-3/2, we transform $|+_{3/2}\rangle$ to any superposition of the basis states $\alpha|+_{3/2}\rangle+\beta|-_{3/2}\rangle$. It will be convenient to write this state, equivalently, as $\lambda|\uparrow\rangle+\kappa|\downarrow\rangle$ for $\lambda=(\alpha+\beta)/\sqrt{2}$ and $\kappa=(\alpha-\beta)/\sqrt{2}$. Analogous to the initialization of the qubit, we slowly ramp down the magnetic field while ramping up the anisotropy on the second site so that $(\lambda|\uparrow\rangle+\kappa|\downarrow\rangle)|\uparrow\uparrow\rangle\rightarrow(\lambda|\uparrow\uparrow\rangle+\kappa|\downarrow\downarrow\rangle)|\rightarrow\rangle$. Applying the same procedure to the third site results in the state  $\lambda|\uparrow\uparrow\uparrow\rangle+\kappa|\downarrow\downarrow\downarrow\rangle=\alpha|+\rangle+\beta|-\rangle$. That is, because we have access to any state on the Bloch sphere defined by $|\pm_{3/2}\rangle$, we have access to any state on the Bloch sphere defined by $|\pm\rangle$.

\subsection{Two-qubit gate}

Consider two spin chains described by the generalized Blume-Capel model in a transverse magnetic field, described by $H$ defined in Sec.~\ref{chain}. As a point of notation, we use tilde to distinguish parameters and operators of the second chain from the first chain. It will be convenient to consider a product state of the two chains, $|\Uparrow\rangle|\tilde\Uparrow\rangle=(|+\rangle|+|-\rangle)(|\tilde+\rangle+|\tilde-\rangle)/2$ which is related to the parity eignenstates by a $\pi/2$-rotation around the $y$ axis. 

Similar to the single qubit operations, we ramp up the transverse magnetic field on all but the first spin sites to decouple the former from the remainder of the chain. This leaves the system described by two copies of Eq.~(\ref{Ha}) with $d^a=\tilde d^a=0$ for $a\neq0$ and $d^5=\tilde d^5$ in general not zero and in the product state $|\uparrow\rangle|\tilde\uparrow\rangle$.  In order to couple the two spins, we use the interaction between the first two sites given by Eq.~(\ref{ising}) and the protocol for the spin-3/2 qubit to apply the Ising XX gate on two spin-3/2 states so that $|\uparrow\rangle|\tilde\uparrow\rangle\rightarrow\exp(-i\Omega_\textrm{I}/2)|\uparrow\rangle|\tilde\uparrow\rangle$. Ramping down the magnetic field on sites $i>1$ and ramping up the anisotropy, $d^5$, in both chains, the state transforms to $\exp(-i\Omega_\textrm{I}/2)|\Uparrow\rangle|\tilde\Uparrow\rangle$. Similar to the transformation made between Eqs.~(\ref{zz})~and~(\ref{xx}), this is an Ising XX gate on the basis of products states of $|\pm\rangle$ and $|\tilde\pm\rangle$.

\section{Discussion}

\label{disc}


Because the Ising XX gate along with rotations about any axis of the Bloch sphere of individual qubits allows for universal quantum computation \cite{debnathNAT16}, our outlined procedure offers universal quantum control of qubits defined in the parity sectors of the generalized Blume-Capel model. Although one can obtain a similar manipulation of the spin-1/2 Ising chain, the spin-3/2 Blume-Capel model offers of quantum operation by geometric manipulation. 

Alternatively, some operations can be performed by pulsing external parameters to apply quantum gates. In particular, consider one chain in the state $|\Uparrow\rangle$ whose first site is in a controllable magnetic field along the Ising axis, $H_B=B S^z_1$. By pulsing this magnetic field for a time $t_Z$, the initial state transforms as $|\Uparrow\rangle\rightarrow\exp(i 3 B t_z /2)|\Uparrow\rangle$. Similarly, $|\Downarrow\rangle\rightarrow\exp(-i 3 B t_z /2)|\Downarrow\rangle$ so that this is a rotation about the $z$ axis of the Bloch sphere by an angle $3Bt_Z/2$ in the basis of $|\Uparrow\rangle$ and $|\Downarrow\rangle$. In the parity basis, this a rotation about the $x$ axis by the same angle, $3Bt_z/2$.

Likewise, a two-qubit gate can be incorporated by pulsing an Ising-like exchange, $H_\textrm{ex}=-\mathcal J S_1^z\tilde S_1^z$, between the first two spins of two spin-3/2 chains. If the chains are initially in the state $|\Uparrow\rangle|\tilde\Uparrow\rangle$, the product state is transformed to $\exp(i 9\mathcal J t_\textrm{ex}/4)|\Uparrow\rangle|\tilde\Uparrow\rangle$ after pulsing the interaction $H_\textrm{ex}$ for a time $t_\textrm{ex}$. Proceeding analogously in this basis, one can show that such an operation leads to an Ising ZZ gate in the $|\Uparrow\rangle$, $|\Downarrow\rangle$ basis and an Ising XX gate in the $|+\rangle$, $|-\rangle$ basis. 

Although we have restricted ourselves to the spin-3/2 chain, fermionization of the Blume-Capel model can be generalized for larger spins in which $|S|=(2^M-1)/2$ for any natural number $M$ in which $S_z$ can written as $M$ spin-1/2 states:
\begin{equation}
S^z=2^M\sum_{i=1}^M2^{i-1}( \mathbb 1\otimes)^{i-1}\sigma^z(\otimes\mathbb 1)^{M-i}\,.
\end{equation}
Consequently, the Blume-Capel model described by Eq.~(\ref{bcm}) generalizes to ladder with $M$ legs. For a chain of $N$ sites there are equivalently $M\times N$ spin-1/2 sites. Notice that because $S_z$ is linear in Pauli matrices, the product of any two spin operators, on either the same site or adjacent sites, will be quadratic in spin-1/2 operators on differing sites, i.e. of the form $\sigma_z^i\sigma_z^j$ for $i\neq j$, which commutes with both $\sigma^1_z$ and $\mathcal P \sigma_z^{NM}$ using the generalization $\mathcal P=\prod_{j=1}^{NM}(-\sigma_x^j)$. Thus, we find the persistence of MBSs at zero energy for higher spin generalizations of the Blume-Capel model which satisfy $|S|=(2^M-1)/2$. Defining a complex fermion analogous to the previous section, we can similarly define the ground states $|\pm\rangle=|\Uparrow\rangle\pm|\Downarrow\rangle$. Moreover, any anisotropy of even order in $S^z$ preserves the energy of the MBSs. As a direct result, a single large spin with anisotropy even in $S^z$ will support two zero energy MBSs. Although generalization the braiding of these MBSs and subsequent geometric manipulation of the quantum states is beyond the scope of our current analysis, should a single spin support quantum operation, an analogous manipulation of the chain parameters should naturally endow a universal set of quantum operation in the parity basis of the chain.

\section{Conclusions}
\label{conc}

In consideration of the Blume-Capel model, we have found that some higher spin chains are capable of supporting MBSs whose occupation naturally defines the basis for a qubit. Guided by the braiding of MBSs, we developed a protocol for universal quantum computation of these qubits by holonomic quantum computing. Having identified spin-3/2 states with MBSs, the vast literature of MBSs can now be directly applied to single spin-3/2 states and their chains. Specifically, methods to encode or read MBSs and apply error correction can now be mapped into the spin-3/2 description. We anticipate this to guide the utilization of higher spin chains for hosting quantum information.

This work was supported as part of the Center for Molecular Magnetic Quantum Materials, 
an Energy Frontier Research Center funded by the U.S. Department of Energy, Office of Science, Basic Energy Sciences under Award No. DE-SC0019330. 
Computations were done using the utilities of National Energy Research Scientific Computing Center and University of Florida Research Computing systems.

\end{document}